\documentclass[twocolumn,showpacs,preprintnumbers,amsmath,amssymb]{revtex4}

\usepackage{graphicx}
\usepackage{bm}

\begin{document}

\title{Casimir momentum of magneto-chiral matter}
\author{James Babington and Bart A. van Tiggelen}
\email{james.babington@grenoble.cnrs.fr}
\affiliation {Laboratoire de Physique et Mod\'{e}lisation des Milieux
Condens\'{e}s,\\ Universit\'{e} Joseph Fourier and CNRS, Maison des
Magist\`eres, 38042 Grenoble, France.}

\date{\today}
\begin{abstract}
We consider a scattering formulation of radiative momentum transfer to a Magneto-Chiral system that is subjected to a constant background magnetic field. The system takes the form of a collection magnetic dipoles that exhibit the Faraday effect. It is shown that the first non-trivial contribution of the momentum transfer to the object from the radiation field occurs at fourth order in the Born series.
\end{abstract}

\maketitle

Casimir and dispersion energies~\cite{Casimir:1948dh,Lifshitz1961} as originally derived in the context of two parallel conducting plates or dielectric bodies, are by now a well understood facet of quantum electrodynamics with measurable experimental implications. The ramifications stretch a broad range of science, from the presence of a cosmological constant~\cite{Elizalde:2007zza}, to stiction forces in nano-scale architecture and applications~\cite{PhysRevE.75.040103}. See~\cite{advancescasimir} and~\cite{RevModPhys.81.1827} for recent surveys.

In the above scenarios, one usually calculates an energy or a force starting from the energy-momentum tensor of electromagnetism. Of course the measurable observable, that is the energy obtained by integration over all space, is but one component of a full tensor of observables. In particular, the time-space components are the associated momentum density of the fields which upon integration over all space gives the total momentum contained in the field~\cite{Kaku:1993ym}. In the case of a single parallel plate, it is precisely because the incident vacuum photons from the left and right tend to cancel that the plate does not develop a net momentum. This preserves the translational invariance of the vacuum. A natural question then to ask is how to find a system whereby a momentum can be displayed.

Magneto-Chiral (MC) systems are bodies that display electromagnetic response functions that depend linearly on an external magnetic field~\cite{vtiggelen02,vtiggelen03,PhysRevLett.104.163901}. In addition to this, the object is also chiral (breaks parity symmetry). This could be because the object as a whole has no reflection symmetry, or because the electromagnetic response function has spatial dispersion. With much of this structure in mind, there have been recent studies~\cite{Feigel:2004zz,Rikken2005298,vanTiggelen:2006zz,PhysRevE.76.066605,Kawka:2010zz,pfeifer2007,vtiggelen08} to see if momentum transfer from the vacuum to bodies with complex media attributes and subjected to external electromagnetic fields is possible. 
The general conclusion seems to be that a momentum can be transferred to a body, though the relative magnitudes of the momenta and assumptions appear not to be consistent with one another.

The main result of this letter is that we demonstrate for a collection of magnetic dipole scatterers that receive a classical perturbative correction to its permeability from an external magnetic field, a non-zero momentum transfer to the body as a whole. It is a fourth order perturbative result and requires at least four scattering centres to be present and held in a rigid configuration. Further, they should be arranged so that the resulting tetrahedron (with the four particles placed at the vertices) has no parity symmetry so that the vacuum photons get to see a chiral structure. If these conditions are fulfilled then a non-zero momentum develops which scales as the fourteenth inverse power of the length scale of the tetrahedron. Whilst the numerical value of this momentum for the pure quantum vacuum case is far too small to be measured experimentally, there are hopes of being able to measure its classical counterpart. The calculations are performed using the notation and conventions in~\cite{scheel-2008-58} and multiple scattering theory.
\begin{figure}[ht]
\begin{center}
\includegraphics[width=4.5cm]{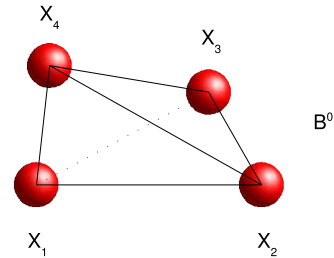}
\caption{The Magneto-Chiral object consists of four point particles in a background complex dielectric, that have a magnetic polarisability $\alpha^{M} (\omega)$ located at positions $\underline{X}_1,\cdots , \underline{X}_4$. The coordinates of the point particles form the vertices of a tetrahedron $\mathcal{T}$. This is subjected to a constant external magnetic field $\textbf{B}^0$.}\label{fig:chiralobject} 
\end{center}
\end{figure}

\paragraph{Radiative momentum transfer formulation.}
\label{sec:formulation}

The system that we investigate is illustrated in Figure~\ref{fig:chiralobject}. It consists of four magnetic dipoles held rigidly in position with respect to one another that is placed in a constant external magnetic field (that has been switched on adiabatically) and a background dielectric that is a complex function of frequency. The question we now try to answer is: \textit{Will a radiative contribution to the momentum of the body result, in particular for the quantum vacuum?}

In order to evaluate the quantum electromagnetic field momentum for the proposed setup, we need to employ a theory that deals with the situation of light-matter interactions in a consistent way. The framework in which we work is that of macroscopic quantum electrodynamics (see for instance~\cite{scheel-2008-58}), which ensures in particular that the equal-time commutation relations are maintained. One could also envisage performing the calculation within a relativistic QFT setting if the potentials can be generalised to appropriate scalar fields. The field momentum derived from the classical Lorentz force law is (at some instant)
\begin{equation}
\label{eq:totalmomenta}
M\bm{v}+\mathbf{P}(t)=constant,
\end{equation}
with
\begin{equation}
\label{eq:emmomenta}
\mathbf{P}(t)=\int d^3 x (\mathbf{E}(t,x)\wedge \mathbf{B}(t,x)).
\end{equation}
We now pass to the quantum version of Equation~(\ref{eq:emmomenta})
\begin{equation}
\label{eq:emmomenta1}
\langle \mathbf{P}(t) \rangle=\int d^3 x \langle 0 |\mathbf{E}(t,x)\wedge \mathbf{B}(t,x)|0\rangle,
\end{equation}
with the spectral decomposition
\begin{eqnarray}
\mathbf{E}(t,x)=  \int^{\infty}_{0}d\omega \mathbf{ E}(\omega, x) e^{-i\omega t}+H.c.,
\end{eqnarray}
and similarly for $\mathbf{B}(t,x)$. They are linked by the Faraday equation
\begin{equation}
\label{eq:Maxwell}
\mathbf{B}(x,\omega) =-\frac{i}{\omega} \nabla \wedge \mathbf{E}(x,\omega),
\end{equation}
together with its hermitian conjugate. By making the appropriate substitutions and point splitting the arguments this can be recast as
\begin{widetext}
\begin{eqnarray}
\label{eq:emmomenta2}
\langle P_a \rangle= \int^{\infty}_{0}d\omega \int^{\infty}_{0}d\omega^{\prime} \int d^3 x \lim_{x^{\prime}\rightarrow x}(-i/\omega) 
( \langle 0 |\mathbf{E}^{\dag}_b(x^{\prime},\omega^{\prime}) \partial^{x}_{a}\mathbf{E}_b(x,\omega)|0\rangle 
- \langle 0 |\mathbf{E}_b^{\dag}(x^{\prime},\omega^{\prime}) \partial_{b}\mathbf{E}_a(x,\omega)|0\rangle ) +c.c.
\end{eqnarray}
\end{widetext}

An important point to appreciate here is that the permeability has both a reciprocal part and a non-reciprocal part due to the external magnetic field. Although the non-reciprocal parts arise from the point interaction where the external magnetic field couples to the scattering object, this is enough to require the correlation functions and thereby the Green tensors, to take a different form. Note here that we are choosing to work with the momentum density associated with the canonical energy momentum tensor rather than the Poynting vector; the latter is expected to integrate to zero~\cite{vtiggelen03} so a permeability is required to distinguish the two vectors. Using the expectation values of the noise currents at zero temperature, we find for the two point function of the physical fields:-
\begin{eqnarray}
\label{eq:fdtwopoint}
 \langle 0 |\mathbf{E}_a(x^{\prime},\omega^{\prime})\mathbf{E}^{\dag}_b(x,\omega)|0\rangle = \frac{\hbar \omega^2}{2i\pi c^2\epsilon_0}\delta(\omega^{\prime}-\omega) \nonumber \\
\times(\mathbf{G}_{ab}(x^{\prime},x|\omega) 
  - \mathbf{G}_{ba}(x,x^{\prime}|\omega)^{*}), 
\end{eqnarray}
with $\langle 0 |\mathbf{E}^{\dagger}_a(x^{\prime},\omega^{\prime})\mathbf{E}_b(x,\omega)|0\rangle =0$. Of course for a standard reciprocal permittivity, the above simplifies to the imaginary part of the Green tensor. 

\paragraph{The Green tensor multiple scattering expansion.}
\label{sec:green}
It remains now to evaluate the Green tensor. This can be done using a multiple scattering expansion. The full Green tensor satisfies the vector Helmholtz equation
\begin{eqnarray}
(\nabla \wedge \bm{\mu}^{-1}(x)\cdot \nabla \wedge \mathbf{G}(x,y|\omega))_{ab}-\frac{\omega^2}{c^2}(\bm{\epsilon}\cdot\mathbf{G})_{ab}(x,y|\omega) \nonumber \\
=\bm{\delta}_{ab}\delta ^3(x-y),
\end{eqnarray}
where we choose $\bm{\epsilon}=\bm{\epsilon}_{ab}(x,\omega)=\bm{\delta}_{ab}$ for the background relative permittivity, whilst the relative permeability of the MC object $\bm{\mu}_{ab}(x,\omega)$ must now account for the magnetic scatterers. Each scatterer has the form
$ \mu_{ab}(x,\omega)-\delta_{ab}=\delta^3(x-x^I)[\alpha^{M}(\omega)\delta_{ab} 
+i\omega\beta^{M}(\omega) \epsilon_{abc}\textbf{B}_{c}^{0}]$, where $\alpha^{M}(\omega)$ and $\beta^{M}(\omega)$ are the magnetic polarisability of the atomic system and a correction due to the external magnetic field, whilst $x^I$ is the position of the $I$-th particle. The magnetic polarisability is given by the expression $\alpha^{M}(\omega)=\alpha^{M}(0)\omega^2_0/(\omega_0^2-\omega^2+i\Gamma \omega)$. By use of the Lorentz force law as applied to a bound system of charges that is overall charge neutral, one finds the relation $\beta^{M}(\omega)=\alpha^{M}(\omega)^2/e$. A further remark about the dimensions of the various quantities is that the combination $|\textbf{B}^{0}| \alpha^{M}(\omega) \omega /e$ is dimensionless. The relative permeability can be turned into the corresponding T-matrix~\cite{vtiggelen02}, together with the multiple scattering expansion of the Green tensor (in symbolic form) $G=G^0+G^0TG^0+G^0TG^0TG^0+\cdots$, where $G^0$ is the free vacuum Green tensor. We shall be specifically interested in the fourth order term corresponding to the scattering from all four particles. An example set of scattering events from $y$ to $x$ contributing to the fourth order Green tensor  is $ G^0(x,X^1)TG^0(X^1,X^2)T\linebreak G^0(X^2,X^3)TG^0(X^3,X^4)TG^0(X^4,y)$.

As a final step in evaluating Equation~(\ref{eq:emmomenta2}) we Fourier transform to momentum space 
\begin{eqnarray}
\label{eq:emmomenta5}
 \langle P_a \rangle &=& \left(\frac{\hbar}{2\pi c^2}\right)\int^{\infty}_{0}d\omega \omega \int \frac{d^3 k}{(2\pi)^3} 
 ik_a \nonumber \\
 &\times &[\mathbf{G}_{bb}(k,k|\omega , \textbf{B}^0) -\mathbf{G}_{bb}(k,k|\omega , \textbf{B}^0)^{*}]. 
\end{eqnarray}
A notable feature of the above expression is the absence of a transverse contribution. The Green tensor can also be split into two distinct pieces. One is the standard reciprocal part due to multiple scatterings from the reciprocal potentials $\alpha^M$ and a second piece that is non-reciprocal and proportional to $\textbf{B}^0$. The reciprocal piece does not contribute to the momentum. 

One is typically interested in evaluating a rotationally averaged version of this momentum, $\bar{P}_a$, particularly from an experimental stand point. A convenient way to do so mathematically is to form the scalar product $\langle P_a \rangle \mathbf{B}^{0}_a$ and then extract from this the bilinear term $\mathbf{B}^{0}_a\mathbf{B}^{0}_b$ using Equation~(\ref{eq:emmomenta5}). From here the replacement $\mathbf{B}^{0}_a\mathbf{B}^{0}_b \rightarrow 1/3(\mathbf{B}^0)^2\delta_{ab}$ is then made. The rotationally averaged momentum then takes the form
\begin{equation}
\bar{P}_a=g \mathbf{B}_a^{0},
\end{equation}
where $g$ is a pseudo-scalar. In the following section, we will evaluate $g=|\bar{P}_a|/|\mathbf{B}^0_a |$. One should also compare this with the result found in~\cite{Rikken2005298} for chiral molecules in a magnetic field.
\begin{figure}[htbp]
\centering
\begin{center}
\includegraphics[width=8.5cm]{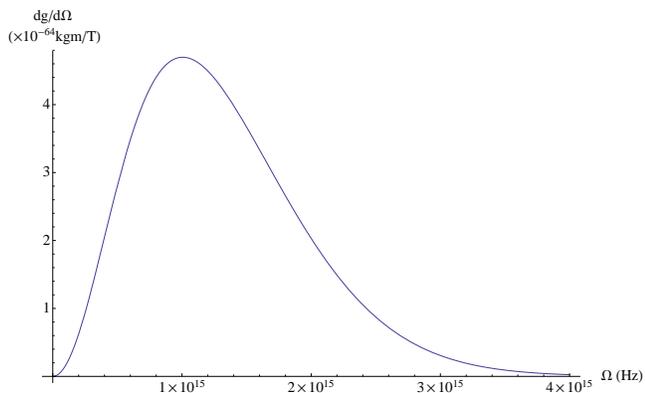}
\caption{A plot of the spectral momentum density (in units of $kgmT^{-1}$) as function of imaginary frequency (Hz) for a collection of four Sodium atoms with parameters $L=100nm$, $\omega_0=3.1\times 10^{15}s^{-1}$.}
\label{fig:momdensity} 
\end{center}
\end{figure}

\paragraph{Results and magnitude of the interaction.}
\label{sec:results}

At this point it is necessary to evaluate Equation~(\ref{eq:emmomenta5}) either symbolically or numerically. We have used \texttt{Mathematica} for this purpose. Before doing this however we can make  some analytical statements. Firstly there are no one, two, or three body amplitude contributions to Equation~(\ref{eq:emmomenta5}). Essentially this is because after the rotational averaging we are left with one three dimensional Levi-Civita tensor that has the effect of forming a determinant of the particle separation vectors (typically a pseudo-scalar of the form $\underline{X}_{12}\cdot(\underline{X}_{23}\wedge \underline{X}_{34})$). It is only at fourth order that this can be non-zero i.e. it is a necessary condition that three of the separation vectors form a linearly independent set and the tetrahedron $\mathcal{T}$ has a finite volume. This is of course why we have chosen to consider a system of four particles from the outset. A second point is that the integral over all space in Equation~(\ref{eq:emmomenta2}) can be done resulting in a simple product of free space propagators over the edges of $\mathcal{T}$ that form a loop. These in particular occur as the overall phase factors in the final expression as the length of the closed loops due to the compounding of the individual propagators. It necessarily results in a closed loop of propagators, that start and finish on the same scattering site. 

This results in the final form (with $k:=\omega / c$) for the pseudo-scalar density
\begin{eqnarray}
\label{eq:finalmomentum}
\frac{dg}{d\omega} &=&\mathrm{Im} \left(\frac{4 \hbar    }{3e\mu_0}   \sum^{3}_{n=1}k^2 e^{ik L_n} \frac{F_n(ikL)}{L^{11}} 
 (\alpha^M(k c)\mu_0)^5\right), \nonumber \\
\end{eqnarray}
where $L_n$ is the length of one of the three inequivalent closed loops (e.g. scatterings from particles in the order 12341 with $n$ indexing the loop). In addition, $L$ is the characteristic length scale of $\mathcal{T}$ (taken to be the length of the shortest edge), and $ F_n(ikL)$ is a dimensionless polynomial of degree $\leq 8$. We are neglecting recurrent scattering and working to fourth order. Then the pseudo-scalar is given by $g=\int^{\infty}_{0}d\omega (dg/d \omega)$. Equation~(\ref{eq:finalmomentum}) can be explicitly evaluated for a collection of different shaped tetrahedra. Notably it vanishes when $\mathcal{T}$ is chosen to be regular i.e. parity invariant. At low frequency
\begin{equation}
{\left. \frac{dg}{d\omega}\right|}_{\omega \rightarrow 0}\sim  \left(\frac{4 \hbar }{3e\mu_0 L^{11}}\right) \left(\alpha^M(0)\mu_0\right)^5\left(\frac{\omega}{c}\right)^2. 
\end{equation}

An example evaluation for a set of vectors $\underline{X}_1=(0,0,0)$, $\underline{X}_2=(L,0,0)$, $\underline{X}_3=(0,2L,0)$, $\underline{X}_4=(L/2,L/2,3L)$ gives a non-zero result. In Figure~\ref{fig:momdensity} we give a sample plot for Sodium atoms of the spectral momentum density as a function of frequency having performed a Wick rotation of frequency to the imaginary axis.

The frequency integral can be evaluated in the retarded limit (where we replace the polarisabilities with their static values) resulting in
\begin{eqnarray}
\label{eq:finalmomentum2}
g = \left(\frac{4\hbar c  }{3e\mu_0}\right)\left(\alpha^M(0)\mu_0\right)^5\left(\frac{1.4\times 10^{4}}{L^{14}}\right), 
\end{eqnarray}
which with the parameters from Figure~\ref{fig:momdensity} and $|\mathbf{B}_a^0|=10T$ give a momentum $|\bar{P}_a|\approx 3.1\times 10^{-47}kgms^{-1}$ with a corresponding MC body speed of $v\approx 2.1\times 10^{-22} ms^{-1}$.

\begin{figure}[htbp]
\centering
\begin{center}
\includegraphics[width=8.5cm]{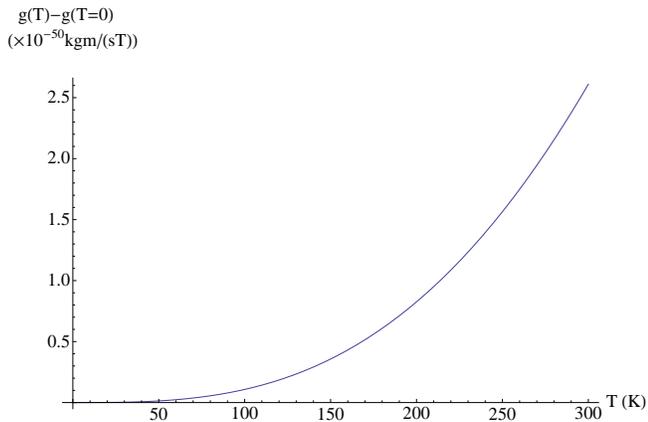}
\caption{A plot of the momentum difference per Tesla as a function of temperature for a collection of four Sodium atoms with parameters as in Figure 2. This is the low temperature regime determined from the length scale $L$ for the first Matsubara frequency. }\label{fig:momentumFT} 
\end{center}
\end{figure}

It is also possible to evaluate the momentum at finite temperature using the standard Matsubara series by inserting a factor of $\coth (\hbar \omega /(2k_BT))$ into Equation~(\ref{eq:finalmomentum}). The Matsubara frequencies are given by $\Omega_m =2\pi k_BTm/\hbar $ where $m$ is a positive integer. In the low temperature limit this can be evaluated for the first few Matsubara frequencies. This is done in Figure \ref{fig:momentumFT} for temperatures less than 300K.
An order of magnitude estimate is also straightforward: the ratio of $(g(T)-g(0))/g(0)$ is approximately for the first Matsubara frequency
 \begin{eqnarray}
\label{eq:ratio}
\frac{g(T)-g(0)}{g(0)} \approx N_1\left(\frac{\alpha^M(i\Omega_1)}{\alpha^M(0)}\right)^5\left(\frac{Lk_BT}{\hbar c}\right)^3,
\end{eqnarray}
where $N_1$ is a numerical constant arising due to the different frequency integration being performed at zero and finite temperature. Using the data for sodium gives $\approx 0.5\%$ correction at $T=300$K.

The high temperature limit can also be calculated which is defined by $T\gg \hbar c/(k_BL)$. In this case the thermal factor at imaginary frequency is $\cot (\hbar \Omega /2k_BT)\approx 2k_BT/\hbar \Omega$. The frequency integral can be done as before with one less power of frequency resulting in
\begin{eqnarray}
\label{eq:highT}
\lim_{T\gg \hbar c/(k_BL)}g(T)&\approx &\left(\frac{8\pi  k_BT}{\mu_0e cL^{13}}\right) (\alpha^M(0)\mu_0)^5,
\end{eqnarray}
where the $\hbar$'s have cancelled out and we enter a regime of classical linear temperature dependence.

So far the calculation we have performed has been for a quantum system of noise currents that have vacuum expectation values proportional to $\hbar$. One can also form a classical counterpart of Equation~(\ref{eq:finalmomentum}) by considering a classical source. For an isotropic distribution of classical radiation such as from a laser, $(dg/d\omega)d\omega$ must be replaced by $\Delta g= dg/d\omega (F/c)(2\hbar \omega^3/(\pi c^3))^{-1}$, where $F$ is the power per unit area of the laser source; the latter factor is the energy density per unit frequency of the vacuum photons we are replacing. An order of magnitude estimate for the momentum can be obtained by evaluating the expression off-resonance (at $\omega = 1.001\omega_0$) to avoid recurrent scattering effects not incorporated here. We use $|\bar{P}_a |\approx |\mathbf{B}^0_a|(\Delta g|_{\omega = 1.001\omega_0}$.  With values of $F=10^{8}Wm^{-2}$ and $\mathbf{B}^0_a|=10T$, this is found to be $|\bar{P}_a |\approx 10^{-35}  kgms^{-1}$ with a corresponding speed $v\approx 10^{-10} ms^{-1}$. The effect can therefore be greatly enhanced by the use of classical sources as compared to the vacuum case.

\paragraph{Summary and perspective.}
\label{sec:conclusion}

In this letter we have investigated if a simple Magneto-Chiral object can develop a momentum from the quantum vacuum. This effect is allowed by symmetry. In itself, the result is a tiny effect for the quantum case, but nevertheless non-zero for the case of magnetic dipoles. For electric dipoles it vanishes rigourously. Thermal corrections have been calculated in the low temperature limit, as well as a high temperature limit. The real frequency  form has also been investigated in a classical setting and the relative enhancement of momentum found off-resonance due to a classical driving source.

\begin{acknowledgments}
We thank S\'{e}bastien Kawka, Geert Rikken and Felipe Pinheiro for useful discussions. This work was supported by ANR through contract PHOTONIMPULS ANR-09-BLAN-0088-01.

\end{acknowledgments}

\end{document}